# Monolayer $MoSe_2$ as an Attenuator of Spintronic Terahertz Emission

Hatice Nur Koyun[1], Alexej Pashkin[1], Ruslan Salikhov[1], Artur Erbe[1,2], Jürgen Faßbender[1,2], Manfred Helm[1,2], Stephan Winnerl[1], Jürgen Lindner[1]

[1]Helmholtz-Zentrum Dresden-Rossendorf, Bautzner Landstraße 400, 01328 Dresden, Germany

[2]Technische Universität Dresden, 01062 Dresden, Germany

## Abstract

Understanding ultrafast spin-to-charge conversion at ferromagnetic/2D interfaces is critical for developing THz-frequency spintronic devices such as ultrafast memories, sensors, and compact THz sources. We investigate this by studying spintronic THz emission (STE) from Py/Au heterostructures capped with transition metal dichalcogenide (TMD) monolayer $MoSe_2$ using time-domain THz spectroscopy. We observe a pronounced reduction in emitted THz amplitude when a monolayer $MoSe_2$ flake covers Py/Au or Py/Pt emitters compared with uncovered regions. A consistent, approximately fourfold attenuation occurs for both systems, despite an order-of-magnitude larger THz radiation from the bare Py/Pt emitter. These results indicate that the suppression is independent of emitter strength. We attribute the attenuation to THz absorption by free charge carriers and trions that are either present due to charge transfer from the metallic layer or photoexcited by the laser pulse, which can significantly diminish the detectable STE signal. This finding provides an explanation for the discrepancy between weak STE signals from FM/TMD heterostructures and magnetotransport reports of sizable spin-orbit torques in similar systems, and highlights the need to account for photoinduced THz attenuation when interpreting ultrafast spin-to-charge conversion measurements in 2D TMD heterostructures.

Ultrafast control of electron spin angular momentum enables to develop terahertz (THz)-frequency devices with versatile potential for magnetic memories, sensors, and THz sources. Central to these advances are photon-driven ultrafast spin-current generation and the rapid conversion of those spin currents into charge currents at hybrid interfaces composed of ferromagnetic (FM) and nonmagnetic (NM) layers. Spin currents may be launched by femtosecond laser-induced demagnetization of an FM film [1-9] or by spin-orbit-mediated processes such as the spin Hall effect (SHE) and the Rashba–Edelstein effect (REE) under THz excitation [10-13]. Efficient ultrafast spin-to-charge interconversion at FM/NM interfaces is therefore essential for broadband THz emission from spintronic emitters [14-18], for picosecond magnetization switching from spin-orbit torques [19], as well as for rapid electrical readout of magnetization states [20-22].

2D materials, particularly transition metal dichalcogenides (TMDs), provide a highly promising platform for efficient spin-to-charge interconversion. Their strong intrinsic spin-orbit coupling, reduced dimensionality, and broken inversion or other crystal symmetries give rise to sizable spin-valley and spin-momentum locking effects that can enhance interfacial spin-orbit-coupling phenomena [23-29]. In addition, TMDs offer favorable electronic tunability (via gating, doping, or varying the layer number), atomically sharp interfaces, and excellent mechanical flexibility, making them attractive for integrating with ferromagnets to engineer and optimize spin-to-charge conversion at FM/TMDs interfaces [25-27, 30]. Indeed, sizable spin-orbit torques (SOTs) have been observed in FM layers interfaced with monolayer TMDs, indicating efficient spin-to-charge interconversion mediated by the inverse REE [27, 30-37].

Despite promising magnetotransport signatures, *ultrafast* spin-to-charge conversion at FM/TMD interfaces remains insufficiently understood and warrants systematic study. Spintronic THz emission (STE) sensitively probes spin-charge dynamics on picosecond timescales [14-18], and STE has been observed in FM/TMD heterostructures spanning bulk, multilayer, and monolayer TMDs [38-44]. Yet the emitted STE from many FM/TMD stacks is markedly smaller than from benchmark FM/Pt interfaces, suggesting reduced picosecond conversion efficiency. Notably, time-domain THz spectroscopy of FM/TMDs subjects the TMD to intense, ultrafast optical excitation that can modify interfacial spin dynamics and spin-to-charge conversion. For example, Victor et al. [45] reported ferromagnetic resonance in $Y_3Fe_5O_{12}/MoS_2$, showing that interfacial spin coherence (i.e., spin-pumping efficiency) is strongly modulated by optical illumination, with a dependence on photon wavelength and flake area-to-perimeter ratio. Moreover, STE efficiency has been shown to depend on pump photon energy, attributed to the requirement that photoexcited, spin-polarized electrons in the ferromagnet overcome the Schottky barrier to transfer into the monolayer TMD [38].

On the other hand, monolayer TMDs host trions, and photoinduced free carriers that generate substantial pump-induced THz conductivity, which can absorb or scatter the radiated THz field and thereby mask otherwise present efficient spin-to-charge conversion. Optical excitation in 2D TMDs creates not only conduction electrons and valence holes but also tightly bound excitons due to strong Coulomb interactions [46]. Both free carriers and charged excitonic complexes couple efficiently to THz fields, producing transient THz absorption and altered scattering rates [47-54] that can strongly influence the measured STE signal.

To disentangle transient THz transmission effects from genuine spin-to-charge conversion, we study STE in $Py/Au/MoSe_2$ heterostructures ($Py = Ni_{81}Fe_{19}$). The Py layer serves as the pump-driven ultrafast spin-current source, while the Au spacer (i) provides a low-dissipation channel for picosecond spin transport owing to its long spin-diffusion length [5], (ii) acts as the primary spin-to-charge converter and THz emitter via interfacial inverse REE processes at the Py/Au interface, with a negligible bulk spin Hall contribution from Au [3, 17, 55-57], and (iii) protects the ferromagnet from oxidation. The monolayer $MoSe_2$ can then either provide additional interfacial conversion of spin currents reaching the Au/TMD interface (enhancing STE) or introduce pump-induced THz absorption that attenuates emission from the Py/Au interface (reducing the observed STE) depending on which interfacial and photoinduced processes dominate.

For these studies metallic thin films were deposited by magnetron sputtering onto double-sided polished, optically transparent, 1-mm-thick quartz glass substrates. Prior to Py/Au growth, we deposited a thin Ta buffer layer to promote adhesion to the substrate and to produce smoother interfaces [58, 59]. In addition, a Ta layer on the other side of the Py layer can enhance the STE response: its spin Hall angle has the opposite sign to the interfacial spin-to-charge conversion at the Py/Au interface, thereby the spin current with the opposite direction at the Py/Ta produces the charge current in the same direction as in the Au layer increasing the overall THz emission [14]. The sample structure fabricated prior to the monolayer TMD transfer was Ta(1.5 nm)/Py(3 nm)/Au(2 nm), with layer thicknesses given in parentheses.

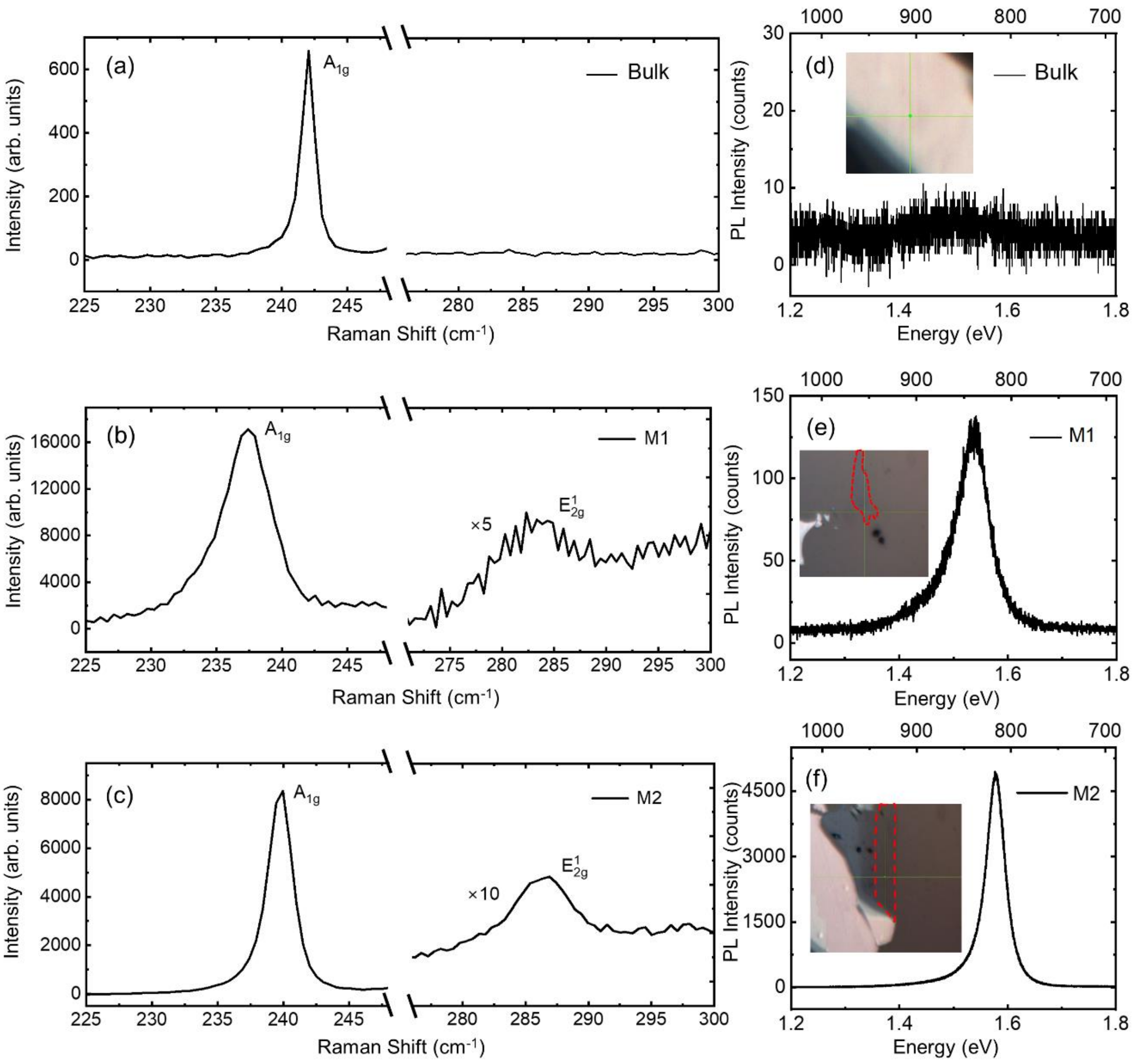

**FIG. 1.** Raman spectra of $MoSe_2$ on Ta(1.5 nm)/Py(3 nm)/Au(2 nm) for (a) bulk, (b) monolayer M1, and (c) monolayer M2 regions after transfer. (d)–(f) Photoluminescence spectra from the corresponding regions. The top scale bar in (d)–(f) indicates photon wavelength; insets in (e) and (f) show the PL probed areas on the sample surface.

Monolayer $MoSe_2$ flakes were mechanically exfoliated from bulk crystals using adhesive "Scotch-tape" onto polydimethylsiloxane (PDMS). Exfoliation was performed by applying normal (downward) pressure to the tape rather than lateral rubbing [60]. Flakes were subsequently transferred onto the multilayer thin films at room temperature. The electronic bandgap of $MoSe_2$ increases from ≈1.3 eV in the bulk to ≈1.58 eV in the monolayer. In the monolayer limit $MoSe_2$, like other TMDs, becomes a direct-bandgap semiconductor with strong light-matter interaction, pronounced excitonic resonances, and detectable optical absorption and emission properties [51, 61]. Optical micrographs and Raman spectra of the exfoliated $MoSe_2$ flakes on PDMS prior to transfer process are shown in supplementary materials Sec. S1 and S2, respectively.

Time-domain THz emission spectroscopy was performed using 800 nm (1.55 eV) laser femtosecond pulses as the optical pump. This photon energy is near the monolayer $MoSe_2$ optical gap, ensuring the absorption that generates free carriers and thus efficiently drives the ultrafast dynamics probed by STE. An amplified Ti:sapphire femtosecond laser system (80 fs pulses, 250 kHz repetition rate) was used. The laser output was split into pump and probe arms:

the pump excites the sample, while the probe beam samples the emitted THz transients. Emitted THz radiation was collected and focused by four off-axis parabolic mirrors onto a 1-mm-thick ZnTe electro-optic crystal for time-domain sampling. The pump beam was focused by a 5x microscope objective (NA = 0.1) to a spot as small as 15 µm on the sample surface. This is an unusual configuration for THz spectroscopy as the small spot size results in strong diffraction of the THz waves and a reduced collection efficiency of the signal-to-noise ratio of the setup. Here, however, we chose this configuration to probe the effect of the small flakes. Details of the experimental setup, including focusing strategies, are provided in supplementary materials Sec. S3.

We identified three flake regions on our sample surface with monolayer flakes of similar widths (≈5-7 µm) and lengths varying from 19 to 43 µm. As a reference, we also measured regions identified as bulk $MoSe_2$ by optical microscopy (see supplementary materials, Sec. S1). The observed sample details are summarized in Table I. Figures 1(b)-1(c) compare Raman spectra from the two $MoSe_2$ monolayer regions (M1, M2) with the bulk region in Fig. 1(a). Spectra were acquired using 532 nm excitation light with a spot size of about 1 µm, using a confocal microscope equipped with a ×100 objective. In the monolayer regions the response is dominated by the in-plane $E^1_{2g}$ phonon mode at ≈287 $cm^{-1}$ and the out-of-plane $A_{1g}$ mode at ≈240 $cm^{-1}$, whereas the $E^1_{2g}$ feature vanishes in the bulk region [61]. This behavior of the $E^1_{2g}$ mode is consistent with reduced dimensionality and suggests the monolayer character of the exfoliated flakes, further supported by the absence of the $B_{2g}$ modes in the Raman spectra (see supplementary materials, Sec. S2).

**TABLE I.** Summary of the studied Ta(1.5 nm)/Py(3 nm)/Au(2 nm)/$MoSe_2$ and Ta(1.5 nm)/Py(3 nm)/Pt(2 nm)/$MoSe_2$ samples, including both bulk and monolayer TMDs. The table lists each sample's short name, monolayer flake dimensions, and the corresponding PL peak intensity and energy.

| Sample | Sample name | Flake width [µm] | Flake length [µm] | PL peak intensity [counts] | PL peak energy [eV] |
|---|---|---|---|---|---|
| Ta/Py/Au/bulk $MoSe_2$ | Bulk | - | - | - | - |
| Ta/Py/Au/flake 1 | M1 | 5 | 19 | 130 | 1.53(7) |
| Ta/Py/Au/flake 2 | M2 | 7 | 31 | 4800 | 1.57(8) |
| Ta/Py/Au/flake 3 | M3 | 6 | 43 | 2000 | 1.51(2) |
| Ta/Py/Pt/flake 4 | M4 | 16 | 20 | 130 | 1.54(2) |

To further confirm the monolayer character, we performed photoluminescence (PL) spectroscopy. Bulk $MoSe_2$ is an indirect-bandgap semiconductor and exhibits weak PL, whereas monolayer $MoSe_2$ becomes a direct-bandgap semiconductor with a single and pronounced PL yield. Monolayers therefore show a strong, sharp PL peak (near ≈1.51-1.58 eV depending on strain and doping), whose marked intensity increase and spectral shape provides a clear signature of the monolayer limit [46, 51, 61]. Figures 1(d)-(f) compare the PL spectra of the bulk region with those of M1 and M2, respectively, revealing substantial PL emission from both monolayer flakes due to their fundamental direct bandgap character, further confirming their monolayer origin. The emitted PL photon energies for all studied monolayers (see Fig. 1(e)-(f) and Fig. 3(a)) vary by approximately 0.06 eV (see Table I). This variation can be attributed to differences in strain and charge transfer from the metallic layers, which may also induce slight shifts in the PL peak positions. More pronounced differences, however, are observed in the PL lineshape and peak intensity (summarized in Table I). The significantly reduced PL peak intensity in the M1 sample suggests that the original quality for that sample is not preserved during the transfer process on the Ta/Py/Au sample. This is also evident in the Raman spectra (Fig. 1(b)), which show a redshift of both the $A_{1g}$ and $E^1_{2g}$ peaks after transfer process compared with the spectra on PDMS (see supplementary materials Sec. S2) [62].

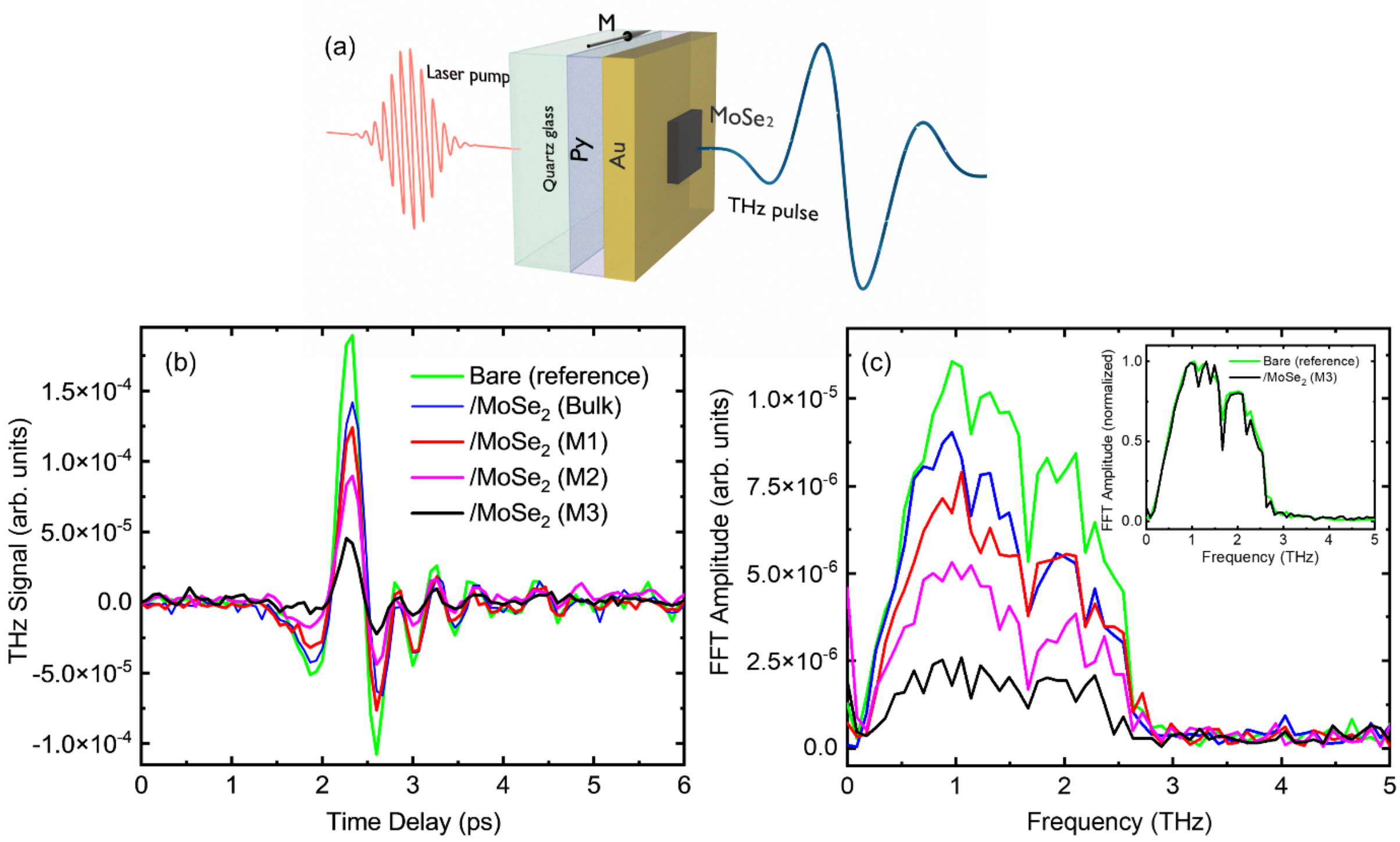

**FIG. 2.** (a) Schematic of the STE experimental geometry. The femtosecond pump pulse is focused from the substrate side onto the TMD flake region; emitted THz radiation is collected from the flake side of the sample. (b) Comparison of STE signals measured on the Ta/Py/Au sample: bare (uncovered) region (green), bulk $MoSe_2$ region (blue), and monolayer flakes M1 (red), M2 (magenta), and M3 (black). The monolayer dimensions and their PL characteristics are listed in Table I. (c) Corresponding Fourier spectra of the waveforms shown in (b) with identical color coding. The inset in (c) compares the normalized spectra between the bare (green line) and M3 flake (black line) regions.

Figure 2(a) shows the STE geometry in our experiment. To avoid absorption of the emitted THz pulses, the focused laser beam was directed through the substrate and incident on the flake from the backside. The Py magnetization (**M**) was saturated with a permanent magnet (~100 mT at the sample). Emitted THz radiation was collected from the flake (air) side using off-axis parabolic mirrors as described above. Time-domain THz signals from different sample regions are presented in Fig. 2(b). It is evident that $MoSe_2$ flakes cause a noticeable reduction in the STE signal compared to the uncovered Py/Au (bare) region, with the THz amplitude reduction being more pronounced in the monolayer flake region than in the $MoSe_2$ bulk. Additionally, we observe that the THz amplitude decreases with increasing monolayer flake size. This trend may be partly attributed to improved overlap between the pump laser beam and larger flakes, leading to stronger attenuation. However, because the largest variation among our samples is in flake length rather than width, and the 15 μm beam spot size is smaller than the length of all flakes, beam-flake overlap alone cannot account for the reduced THz emission. We therefore propose that structural quality is also a key factor, as THz attenuation is strongest in the M2 and M3 regions, which exhibit PL intensities an order of magnitude higher than those of the M1 flake region (Table I). The higher structural quality, evidenced by the stronger PL emission, suggests increased carrier mobility in this layer, which in turn leads to enhanced THz absorption. The fast Fourier transforms (FFTs) in Fig. 2(c) show no appreciable change in spectral shape (see inset in Fig. 2(c)), indicating that the monolayer-induced THz attenuation is approximately frequency-independent over the measured range.

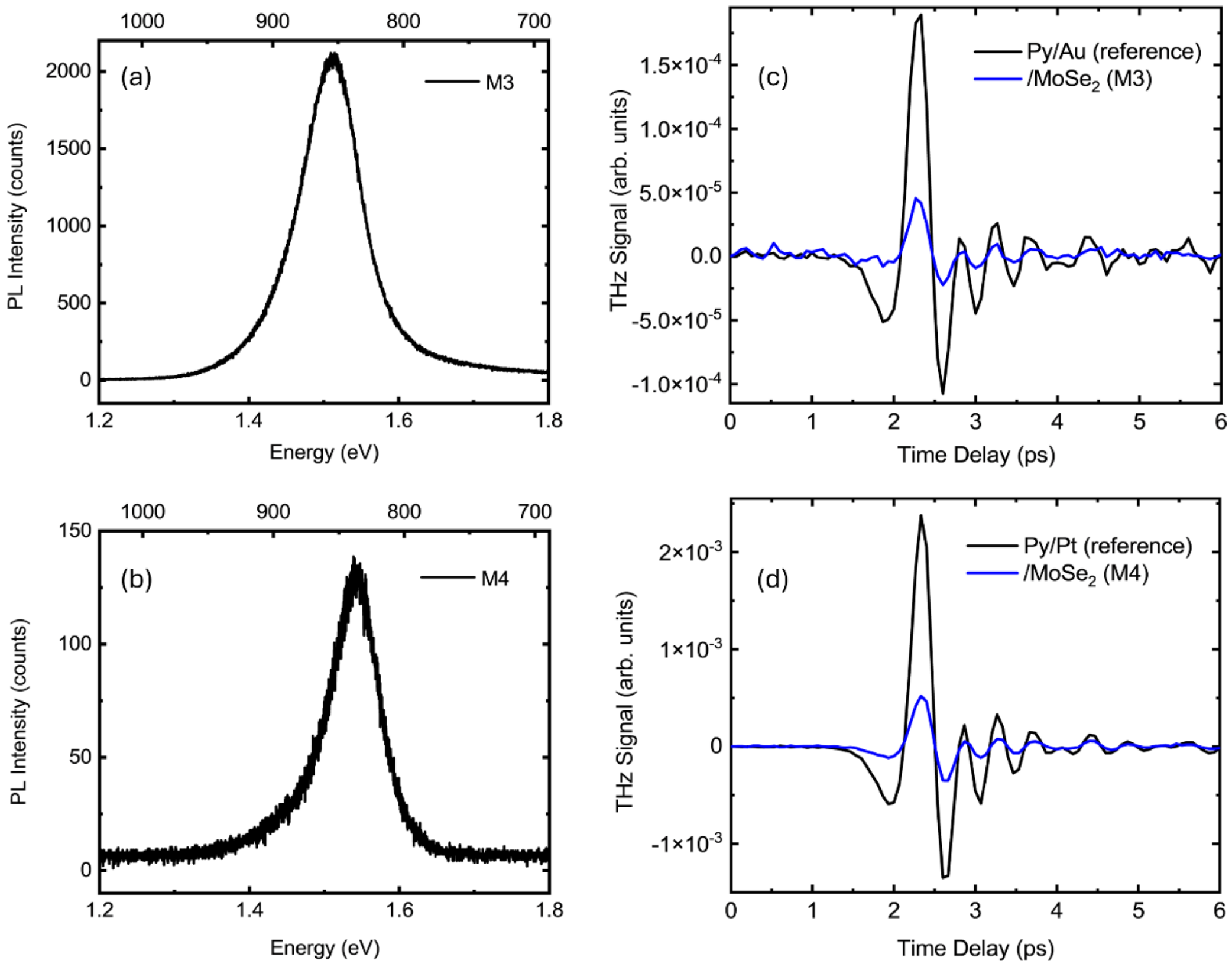

**FIG. 3.** (a)–(b) Photoluminescence spectra of monolayer $MoSe_2$ transferred onto Ta(1.5 nm)/Py(3 nm)/Au(2 nm) and Ta(1.5 nm)/Py(3 nm)/Pt(2 nm), respectively (samples M3 and M4 in Table I). (c)–(d) Corresponding STE signals measured on bare (black) and monolayer (blue) regions.

Since $MoSe_2$ monolayers on Py/Au resulted in substantial THz attenuation rather than enhancement, we tested their effect on a Py/Pt sample, for which STE amplitude is expected to be an order of magnitude larger [14-17]. Figures 3(a)-3(b) compare the PL from flakes on Py/Au (M3) and Py/Pt (M4). The PL spectral shape and peak energy are essentially similar, suggesting that the M4 flake is a $MoSe_2$ monolayer. Figures 3(c)-3(d) compare the STE signals: the uncovered Py/Pt region exhibits a THz peak amplitude an order of magnitude larger than the uncovered Py/Au reference, yet the monolayer region on Py/Pt is attenuated by factor of 4 - the same factor observed for the Py/Au sample. This implies that the monolayer transmits approximately 25% of the THz amplitude regardless of the emitter strength, supporting that Drude absorption and scattering in the single-layer $MoSe_2$ are responsible for the THz attenuation.

The 75% reduction in the THz peak amplitude observed over the monolayer region is surprisingly large. Our model of THz transmission through a highly conducting 2-nm-thin Pt film predicts a 65% reduction [22], i.e., less attenuation than measured for $MoSe_2$ monolayers.

In general, reported conductivities of 2D TMDs are typically at least an order of magnitude lower than that of Pt [37, 54], suggesting that intrinsic THz transmission through an isolated monolayer should be substantially higher than for Pt films. While charge transfer from adjacent Au or Pt layers can markedly alter the electronic properties of $MoSe_2$ [63] and thus enhance THz absorption and scattering, we do not expect this interfacial doping to be the sole cause of the large attenuation.

The much smaller attenuation observed for bulk areas of the flake (Fig. 2b) indicates that photoexcitations associated with the direct-gap monolayer response make a substantial contribution. To quantify this, we estimate a monolayer $MoSe_2$ sheet photoconductivity of ~0.06 S/sq at a pump fluence of 17 mJ/cm², which exceeds the ~0.024 S/sq needed to produce a 75% reduction in THz amplitude (see supplementary materials, Sec. S4). This readily accounts for the observed STE suppression, even considering that the pump reaching the TMD is reduced by transmission losses in the metallic films. Moreover, charge transfer from adjacent Au or Pt can dope the $MoSe_2$, increasing carrier density and further enhancing THz absorption, thereby strengthening the observed attenuation. These trends are consistent with reports of strong pump-induced THz photoconductivity in monolayer TMDs [54].

While photoinduced THz absorption plausibly explains the suppression, a spintronic contribution could, in principle, also reduce the STE via destructive interference. In particular, a negative spin Hall conductivity in $MoSe_2$ (reported for $WSe_2$ and inferred to be similar in $MoSe_2$ [40, 64]) would invert the STE waveform (180° phase shift) relative to emission from the Py/Au or Py/Pt interfaces, lowering the peak amplitude. We consider this effect negligible for our data: the attenuation is essentially identical for Py/Au/$MoSe_2$ and Py/Pt/$MoSe_2$ (Fig. 3c-d), even though bare Py/Pt emits approximately 10-fold larger THz amplitude than Py/Au, and Pt's much shorter spin-diffusion length makes a sizable additional STE contribution at the Pt/$MoSe_2$ interface improbable. We further note that systematic separation of such spintronic interference from absorption is impractical here because the measured THz amplitude is highly sensitive to flake size and quality (see M1-M3 on Py/Au). The monolayer on Pt (M4) shows PL peak intensity comparable to M1 on Py/Au, indicating similar strain/disorder, yet M4 exhibits the same ~75% STE reduction as the higher-quality M3. We attribute this primarily to geometric overlap: the M4 flake width exceeds the pump-beam diameter, ensuring complete illumination - unlike all flakes on Py/Au (see Table I).

To summarize, time-domain THz spectroscopy reveals a pronounced (at least four-fold) reduction in STE amplitude from Py/Au and Py/Pt emitters when covered by a $MoSe_2$ monolayer flake. We attribute this reduction to enhanced transient THz absorption and/or scattering by electrons transferred from the metal and by photoexcited carriers. This attenuation can mask the intrinsic ultrafast spin-to-charge conversion at FM/TMD interfaces, producing an apparent decrease in conversion efficiency and offering a plausible explanation for the discrepancy with magnetotransport studies, which typically report sizable spintronic effects in 2D TMDs. To probe the intrinsic ultrafast spin-to-charge conversion at FM/TMD interfaces, pump illumination of the TMD must be avoided. One straightforward approach is to use a metallic spacer thicker than the optical penetration depth (≈10-20 nm) to block direct excitation of the TMD. However, such increased metal thickness typically suppresses the emitted THz amplitude [14]. An alternative is to induce FM demagnetization with sub-bandgap THz excitation (photon energy well below the TMD bandgap), thereby driving spin currents without directly photoexciting the monolayer [8, 21]. Because THz beams are millimeter-scale, these schemes require large-area, continuous 2D films (e.g., epitaxially grown TMDs or recently developed continuous TMD heterostructures) to enable reliable, spatially uniform measurements.

**Acknowledgments**

The authors are grateful to Thomas Naumann for experimental and technical support. R.S. acknowledges support by the Deutsche Forschungsgemeinschaft (DFG, German Research Foundation) – project number 566522216.

**Competing interests**

The authors have no conflicts to disclose.